\documentclass[12pt,draftcls,onecolumn]{IEEEtran}
\usepackage{setspace}
\usepackage{clrscode}
\usepackage{pict2e}

\usepackage{amsmath}
\usepackage{amssymb}
\usepackage[dvips]{graphicx}
\usepackage{epsfig}
\usepackage{hyperref}
\usepackage{algorithm}
\usepackage{algorithmic}
\usepackage{caption}
\usepackage{subcaption}
\usepackage{capt-of}

\usepackage{color}

\usepackage[ps2pdf,
bookmarks=false,
bookmarksnumbered=false, 
bookmarksopen=false, 
colorlinks=false]{}

\newcommand{\rme}{{\mathrm{e}}}
\newcommand{\E}{\mathbb{E}}

\newcommand{\mI}{\mathcal{I}}
\newcommand{\mx}{\mathcal{X}}

\newcommand{\hn}{\hat{n}}

\makeatletter

\newcommand{\Rmnum}[1]{\expandafter\@slowromancap\romannumeral #1@}
\makeatother

\newcommand{\tSNR}{\text{\footnotesize{SNR}}}

\newcommand{\ben}{\frac{E_b}{N_0}}

\newtheorem{Lem}{Theorem}
\newtheorem{Rem}{Remark}

\newtheorem{Prop}{Proposition}

\begin{document}

%
\title{Optimal Power Control for Fading Channels with Arbitrary Input Distributions and Delay-Sensitive Traffic}

\author{
\IEEEauthorblockN{Gozde Ozcan and M. Cenk Gursoy}
\thanks{Gozde Ozcan was with the Department of Electrical
Engineering and Computer Science, Syracuse University, and she is now with Broadcom Limited, Irvine, CA. M. Cenk Gursoy is with the Department of Electrical
Engineering and Computer Science, Syracuse University, Syracuse, NY, 13244
(e-mail: gozcan@syr.edu, mcgursoy@syr.edu).}}

\maketitle
\thispagestyle{empty}

\begin{spacing}{1.75}

\begin{abstract}
This paper presents the optimal power control policies maximizing the effective capacity achieved with arbitrary input distributions subject to an average power constraint and quality of service (QoS) requirements. The analysis leads to simplified expressions for the optimal power control strategies in the low power regime and two limiting cases, i.e., extremely stringent QoS constraints and vanishing QoS constraints. In the low power regime, the energy efficiency (EE) performance with the constant-power scheme is also determined by characterizing both the minimum energy per bit and wideband slope for arbitrary input signaling and general fading distributions. Subsequently, the results are specialized to Nakagami-$m$ and Rician fading channels. Also, tradeoff between the effective capacity and EE is studied by determining the optimal power control scheme that maximizes the effective capacity subject to constraints on the minimum required EE and average transmission power. Circuit power consumption is explicitly considered in the EE formulation. Through numerical results, the performance comparison between constant-power and optimal power control schemes for different signal constellations and Gaussian signals is carried out. The impact of QoS constraints, input distributions, fading severity, and average transmit power level on the proposed power control schemes, maximum achievable effective capacity and EE is evaluated.
\end{abstract}

\begin{IEEEkeywords}
Effective capacity, energy efficiency, fading channel, low-power regime, mutual information, MMSE, optimal power control, QoS constraints.
\end{IEEEkeywords}


\section{Introduction}
Transmission power is one of the key factors in wireless communications since it is not only a limited resource but it also accounts for the significant portion of the total power consumption. Hence, a common type of resource adaptation is to efficiently vary the transmission power over time as a function of the channel conditions in order to enhance the system performance. A legion of studies has been conducted on power adaptation in wireless systems. It is well known that \emph{water-filling} algorithm is the optimal power control policy, maximizing the spectral efficiency when the input is Gaussian distributed and perfect channel side information (CSI) is available at the transmitter \cite{goldsmith}. On the other hand, Shannon capacity does not address quality of service (QoS) constraints in the form of constraints on buffer overflow probabilities or queueing delays.

Many important wireless applications (e.g. mobile streaming/interactive video, voice over IP (VoIP), interactive gaming and mobile TV) require certain QoS guarantees for acceptable performance levels at the end-user. In \cite{dapeng}, effective capacity is proposed to serve as a suitable metric to quantify the performance of wireless systems under statistical QoS constraints. In particular, effective capacity provides the maximum throughput in the presence of limitations on the buffer-overflow/delay-violation probabilities by capturing the asymptotic decay-rate of the buffer occupancy.

The analysis and application of effective capacity in wireless systems have attracted growing interest in recent years. For instance, the authors in \cite{tang} first proposed the optimal power and rate adaptation schemes that maximize the effective capacity of a point-to-point wireless communication link. Then, they considered multichannel communications and derived the optimal power control policy for multicarrier and multiple-input and multiple-output (MIMO) systems in \cite{tang2}. The authors \cite{helmy} formulated the energy efficiency (EE) by the ratio of effective capacity to the total power consumption including circuit power, and determined the optimal power allocation for multicarrier systems over a frequency-selective fading channel. The work in \cite{liu} mainly focused on energy-efficient power allocation for delay-sensitive multimedia traffic in both low- and high-signal-to-noise ratio (SNR) regimes. Recently, the authors in \cite{musavian} determined the QoS-driven optimal power control policy in closed-form to maximize the effective capacity subject to a minimum required EE level. The authors in \cite{gursoy} employed the notion of effective capacity and analyzed the EE under QoS constraints in the low-power and wideband regimes by characterizing the minimum energy per bit and wideband slope. Also, the authors in \cite{ru} derived the minimum energy per bit and the wideband slope region for the dirty paper coding (DPC) and time division multiple access (TDMA) schemes under heterogeneous QoS constraints. Additionally, the authors in \cite{zhong} obtained the effective capacity of correlated multiple-input single-output (MISO) channels and further analyzed the performance in low- and high-SNR regimes. Moreover, the authors in \cite{benkhelifa} derived the asymptotic expression of the effective capacity in the low power regime for a Nakagami-$m$ fading channel.

The common assumption in the aforementioned works was that the input signal is Gaussian distributed. However, it may be difficult to realize Gaussian inputs in practice. Therefore, practical applications generally employ inputs from discrete constellations such as pulse amplitude modulation (PAM), quadrature amplitude modulation (QAM) and phase-shift keying (PSK). Recently, the authors in \cite{lozano} identified the optimal power allocation scheme called \emph{mercury/water-filling} for parallel channels with arbitrary input distributions subject to an average power constraint by using the relation between the mutual information and minimum mean square error (MMSE). Subsequently, in \cite{nguyenm}, a low-complexity, suboptimal power adaptation scheme was proposed in order to minimize the outage probability and maximize the ergodic capacity for block-fading channels with arbitrary inputs subject to peak, average, and peak-to-average power constraints. The work in \cite{gunduz} mainly focused on the power allocation for Gaussian two-way relay channels with arbitrary signaling in the low and high power regimes. Despite recent interest in the performance achieved with arbitrarily distributed input signals, most works have not incorporated QoS considerations into the analysis. Therefore, it is of significant interest to analyze the effective capacity achieved with arbitrarily distributed signals under statistical QoS constraints (imposed as limitations on buffer-overflow/delay-violation probabilities). Recently, we have considered Markovian sources and obtained the optimal power control schemes that maximize the EE of wireless transmissions with finite discrete inputs \cite{ozcan}. Different from that work, we in this paper first derive the optimal power adaptation scheme that maximizes the throughput of delay-sensitive traffic (quantified by the effective capacity) achieved with arbitrary input signaling subject to an average transmit power constraint. Then, we analyze the proposed optimal power policy under extremely stringent QoS constraints and also vanishing QoS constraints. Also, we analyze the performance with arbitrary input signaling in the low power regime by characterizing the minimum energy per bit and wideband slope for general fading distributions. In addition, we provide a simple approximation for the optimal power control policy in the low power regime. Finally, we consider the tradeoff between the effective capacity and EE by formulating the optimization problem to maximize the effective capacity subject to constraints on the minimum required EE and average transmission power.

The rest of the paper is organized as follows: Section \ref{sec:system_model} introduces the system model. Section \ref{sec:prelim} describes the notion of effective capacity. In Section \ref{sec:opt_power_Ec}, the optimal power control policy maximizing the effective capacity achieved with arbitrary input distributions is derived. In Section \ref{sec:opt_power_limiting}, the optimal power control in limiting cases is analyzed. Section \ref{sec:low_power_analysis} provides low-power regime analysis of the effective capacity attained with constant-power scheme and the optimal power control. Before presenting the numerical results in Section \ref{sec:num_results}, the optimal power control that maximizes the effective capacity subject to a minimum EE constraint is obtained in Section \ref{sec:opt_power_EEmin}. Finally, main concluding remarks are provided in Section \ref{sec:conc}.

\section{System Model} \label{sec:system_model}
In this paper, we consider a point-to-point wireless communication link between the transmitter and the receiver over a flat fading channel. Hence, the received signal is given by
\begin{align}
y[i]=h[i]x[i]+n[i] \hspace{0.5cm}i=1,2,\dots
\end{align}
where $x[i]$ and $y[i]$ denote the transmitted and received signals, respectively, and $n[i]$ is a zero-mean, circularly symmetric, complex Gaussian random variable with variance $N_0 B$ where $B$ denotes the bandwidth. It is assumed that noise samples \{$n[i]$\} form an independent and identically distributed (i.i.d.) sequence. Also, $h[i]$ represents the channel fading coefficient,
and the channel power gain is denoted by $z[i]=|h[i]|^2$.

If the transmitter perfectly knows the instantaneous values of \{$h[i]$\}, it can adapt its transmission power according to the channel conditions. Let $P[i]$ denote the power allocated in the $i^{\text{th}}$ symbol duration. Then, the instantenous received signal-to-noise ratio, $\tSNR$ can be expressed as $\gamma= \frac{P[i]z[i]}{N_0B}$. The average transmission power is constrained by $\bar{P}$, i.e., $\E\{P[i]\} \le \bar{P}$, which is equivalent to $\E\{\mu [i]\} \le \tSNR$, where $\mu [i]=\frac{P[i]}{N_0B}$ and $\tSNR=\frac{\bar{P}}{N_0B}$.
In the rest of the analysis, we omit the time index $i$ for notational brevity. We express the transmitted signal $x$ in terms of a normalized unit-power arbitrarily distributed input signal $s$. Now, the received signal can be expressed as
\begin{align}
\hat{y}=\sqrt{\rho}s+\hn,
\end{align}
where $\rho=\mu z$, jointly representing the channel gain and transmission and noise powers, and $\hn$ is the normalized Gaussian noise with unit variance. Let us define the input-output mutual information $\mI (\rho)$ as
\begin{align}
\mI (\rho)= \mI (s; \sqrt{\rho}s+\hn).
\end{align}
For Gaussian input $s$, $\mI (\rho)=\log_2(1+\rho)$, while for any input signal $s$ belonging to a constellation $\mathcal{X}$, we have
\begin{equation}
\begin{split} \label{eq:I_computation}
\mI (\rho) = &\log_2 |\mx |-\frac{1}{\pi |\mx |} \\ &\hspace{-0.5cm}\times \sum_{s \in \mx} \int \log_2 \Bigg(\sum_{s'\in \mx} \rme^{-\rho |s - s'|^2-2\sqrt{\rho}\mathcal{R}\{(s-s')^* \hn\}}\Bigg) \rme^{-|\hn|^2} d\hn,
\end{split}
\end{equation}
where $\mathcal{R}\{\}$ denotes the operator that takes the real part and the integral is evaluated in the complex plane . The relation between the mutual information and the minimum mean-square error (MMSE) is given by \cite{guo}
\begin{align} \label{eq:first_der_mutual}
\dot{\mI}(\rho)= \text{MMSE}(\rho)\log_2\rme,
\end{align}
which is used to derive the power control policy for independent and parallel channels \cite{lozano}. Above,  $\dot{\mI}(.)$ denotes the first derivative of the mutual information, $\mI(\rho)$, with respect to $\rho$. The MMSE estimate of $s$ is given by
\begin{align}
\hat{s}(\hat{y},\rho)=\E\{s \mid \sqrt{\rho}s+\hn\}.
\end{align}
Then, the corresponding MMSE is
\begin{align}
\text{MMSE}(\rho)=\E\{|s-\hat{s}(\hat{y},\rho)|^2\}.
\end{align}
It should be noted that $\text{MMSE}(.) \in [0,1]$. When the input signal $s$ is Gaussian, $\text{MMSE}(\rho)=\frac{1}{1+\rho}$. On the other hand, for any arbitrarily distributed signal $s$ with a constellation $\mathcal{X}$, we have
\begin{align} \label{eq:MMSE_general}
\text{MMSE}(\rho)\!=\!1\!-\!\frac{1}{\pi |\mx |}\int \frac{\big|\sum_{s \in \mx }s\rme^{2\sqrt{\rho} \mathcal{R}\{\hat{y}s^*\}-\rho|s|^2}\big|^2}{\sum_{s\in \mx}\rme^{2\sqrt{\rho}\mathcal{R}\{\hat{y}s^*\}-\rho|s|^2}}\rme^{-|\hat{y}|^2}d\hat{y}.
\end{align}
For a specific constellation such as 4-pulse amplitude modulation (4-PAM), MMSE is given by (\ref{eq:MMSE_4PAM}) on the next page \cite{lozano}.
\begin{figure*}
\begin{align}\label{eq:MMSE_4PAM}
\text{MMSE}^\text{4-PAM}(\rho)&=1-\int_{-\infty}^{\infty}\frac{\Big(\rme^{-8\rho/5}\sinh\big(6\sqrt{\frac{\rho }{5}}\phi\big)+\sinh\big(2\sqrt{\frac{\rho }{5}}\phi\big)\Big)^2}{\rme^{-8\rho/5}\cosh\big(6\sqrt{\frac{\rho }{5}}\phi\big)+\cosh\big(2\sqrt{\frac{\rho }{5}}\phi\big)}\frac{\rme^{-\phi^2-\rho/5}}{10\sqrt{\pi}} d\phi.
\end{align}
\hrule
\end{figure*}
The MMSE for 16-quadrature amplitude modulation (16-QAM) can be readily determined by using the MMSE of 4-PAM in (\ref{eq:MMSE_4PAM}) as follows:
\begin{align}
\text{MMSE}^\text{16-QAM}(\rho)&=\text{MMSE}^\text{4-PAM}\left(\frac{\rho }{2}\right).
\end{align}
As further special cases, binary phase-shift keying (BPSK) (or equivalently 2-PAM), and quadrature phase-shift keying (QPSK) (or equivalently 4-QAM), the above MMSE expressions can be further simplified as follows \cite{lozano}:
\begin{align}
\text{MMSE}^\text{BPSK}(\rho)&=1-\int_{-\infty}^{\infty}\tanh(2\sqrt{\rho}\phi)\frac{\rme^{-(\phi-\sqrt{\rho })^2}}{\sqrt{\pi}}d\phi, \\
\text{MMSE}^\text{QPSK}(\rho)&=\text{MMSE}^\text{BPSK}\left(\frac{\rho }{2}\right).
\end{align}
It should be noted that the mutual information in (\ref{eq:I_computation}) and MMSE expression in (\ref{eq:MMSE_general}) can be easily computed by decomposing them into two dimensional real integrals and applying Gauss-Hermite quadrature rules \cite{abramovitz}.

\section{Preliminaries} \label{sec:prelim}
Before introducing the optimal power adaptation problem aiming at maximizing the effective capacity in the next section, we briefly review the notion of effective capacity. Based on the theory of large deviations, effective capacity identifies the maximum constant arrival rate that can be supported by a time-varying service process while the buffer overflow probability decays exponentially fast asymptotically for large buffer thresholds. More specifically, effective capacity is the maximum constant arrival rate in a queuing system with service process $\{R[j]\}$ such that the \emph{stationary queue length $Q$} satisfies
\begin{align} \label{eq:overflowlimit}
\lim_{q_{\max} \to \infty} -\frac{\log \Pr\{Q \ge q_{\max}\}}{q_{\max}} =\theta
\end{align}
where $q_{\max}$ denotes the buffer overflow threshold, $\Pr\{Q \ge q_{\max}\}$ is the buffer overflow probability, and $\theta$ is called the QoS exponent. For a discrete-time stationary and ergodic service process $\{R[j]\}$, the effective capacity is given by \cite{dapeng} \cite{chang}\footnote{We note that the effective capacity $C_E(\tSNR,\theta)$ is a function of both $\tSNR$ and the QoS exponent $\theta$. However, in order to simplify the notation in the remainder of the paper, we henceforth express the effective capacity explicitly only in terms of $\tSNR$ and denote it by $C_E(\tSNR)$ due to the fact the we generally conduct the analysis and obtain characterizations for given fixed $\theta$.
\\
\indent We further note that $\log$ (without an explicit base) denotes logarithm to the base of $e$ (i.e., the natural logarithm) throughout the text.}:
\begin{gather}
\label{eq:eff-cap-def}
C_E(\tSNR,\theta) = -\lim_{t\rightarrow\infty}\frac{1}{\theta
t}\log{\mathbb{E}\{e^{-\theta \sum_{j=1}^tR[j]}\}}.
\end{gather}
For large $q_{\max}$, the limit in (\ref{eq:overflowlimit}) implies that the buffer overflow probability can be approximated as
\begin{align}
\Pr\{Q  \ge q_{\max}\} \approx
\rme^{-\theta q_{\max}}, \label{eq:bufferconstraint}
\end{align}
where $\theta$ characterizes the exponential decay rate of the buffer overflow probability. From the above approximation, we can see that larger values of $\theta$ indicate more stringent QoS constraints since it imposes faster decay rate. Smaller $\theta$ reflects looser constraints.

In addition, the delay-bound violation probability is characterized to decay exponentially and can be approximated as \cite{du}
\begin{align}
\Pr\{D \geq D_{\text{th}} \} \approx \varphi \rme^{-\theta C_E(\tSNR)D_{\text{th}}},
\end{align}
where \emph{$D$ denotes the steady state queueing delay}, $D_{\text{th}}$ represents the delay threshold, $\varphi = \Pr\{Q > 0\}$ is the probability that the buffer is nonempty, which can be approximated by the ratio of the average arrival rate to the average service rate \cite{chang}.

When the service process $\{R[j]\}$ is i.i.d., the effective capacity simplifies to
\begin{align}
\label{eq:effective_capacity}
C_E(\tSNR)=-\frac{1}{\theta}\log(\E\{e^{-\theta R[j]}\}).
\end{align}

\section{Optimal Power Control} \label{sec:opt_power_Ec}
Our goal is to derive the optimal power control policy that maximizes the effective capacity achieved with an arbitrary input distribution, which can be found by solving the following optimization problem
\begin{align}
\label{eq:EC_opt}
C_E^{\text{opt}} (\tSNR)=&\max_{
\substack{\mu( \theta, z)}} -\frac{1}{\theta TB}\log(\E\big\{\rme^{-\theta TB \mI (\mu (\theta, z)z)}\big\})\\
&\hspace{-.5cm}\text{subject to} \hspace{0.2cm}\E\{\mu (\theta, z)\}  \le \tSNR,
\end{align}
where $T$ is the frame duration and the expectation $\E\{.\}$ is taken with respect to the channel power gain $z$. Above, $C_E^{\text{opt}} (\tSNR)$ denotes the maximum effective capacity attained with the optimal power control scheme and $\mu(\theta, z) = \frac{P(\theta,z)}{N_0B}$ represents the instantaneous transmission power normalized by the noise power, and both $\mu(\cdot)$ and the power $P(\cdot)$ are expressed as functions of the QoS exponent $\theta$ and channel power gain $z$. Moreover, the instantaneous transmission (or equivalently service) rate achieved with an arbitrarily distributed input is formulated as proportional to the input-output mutual information $\mI (\mu (\theta, z)z)$\footnote{We note that in our setting the service process of the buffer, denoted by $R$ in the effective capacity formula in (\ref{eq:effective_capacity}), is the instantaneous transmission rate over the fading channel, which depends on the type of input signal used for transmission. For instance, if the transmitted signal is the capacity-achieving Gaussian signal, then $R  = T B \log_2 \left(1 + \frac{P(\theta,z) z}{N_0 B}\right)$, which is the mutual information achieved by the Gaussian input. For any other input, rate is proportional to the mutual information $\mI$ achieved by this input.}.

We first have the following characterization for the optimal power control policy.

\begin{Lem} \label{teo1}The optimal power control, denoted by $\mu_{\text{opt}}(\theta, z)$, which maximizes the effective capacity in (\ref{eq:EC_opt}), is given by
\begin{align}
\label{eq:opt_power}
\mu_{\text{opt}}(\theta, z) = \begin{cases}
0, &z \le \alpha\\
\mu ^* (\theta, z), &z> \alpha
\end{cases},
\end{align}
where $\mu ^* (\theta, z)$ is solution to
\begin{align} \label{eq:Ec_opt_equation}
\rme^{-\theta TB \mI (\mu ^* (\theta, z)z)}\text{MMSE}(\mu ^* (\theta, z)z)z=\alpha
\end{align}
and $\alpha$ satisfies
\begin{align}\label{eq:alpha_constraint}
\int_{\alpha}^{\infty}\mu ^* (\theta, z)f(z)dz=\tSNR.
\end{align}
Above, $f(z)$ is the probability density function (PDF) of the channel power gain $z$.
\end{Lem}
\emph{Proof:}
The mutual information $\mI (\rho)$ is a concave function of $\rho$ since its second derivative is negative, i.e., $\ddot{\mI}(\rho) = - \E\{(\E\{|s[i]-\hat{s}[i]|^2\big|\hat{y}[i]\})^2+|\E\{(s[i]-\hat{s}[i])^2\big|\hat{y}[i]\}|^2\} < 0$ \cite{payaro}. Subsequently, $-\theta TB \mI (\rho)$ is a convex function for given values of $\theta$, $T$, $B$, and $\rme^{-\theta TB \mI (\mu (\theta, z)z)}$ is a log-convex function of $\mu$, which takes non-negative values. Since expectation preserves log-convexity, $\E\big\{\rme^{-\theta TB \mI (\mu (\theta, z)z)}\big\}$ is also log-convex in $\mu$ \cite{boyd}. This implies that $\log(\E\big\{\rme^{-\theta TB \mI (\mu (\theta, z)z)}\big\})$ is a convex function of $\mu$. Since the negative of a convex function is concave, it follows that the objective function in (\ref{eq:EC_opt}) is concave in $\mu$. Since logarithm is a monotonic increasing function, the optimal power control policy can be found by solving the following minimization problem:
\begin{align}
\label{eq:EC_opt_min}
&\min_{
\substack{\mu( \theta, z)}} \E\big\{\rme^{-\theta TB \mI (\mu (\theta, z)z)}\big\}\\ \label{eq:constraint}
&\text{subject to} \hspace{0.2cm}\E\{\mu (\theta, z)\}  \le \tSNR.
\end{align}
Accordingly, we first write the expectations in (\ref{eq:EC_opt_min}) and (\ref{eq:constraint}) as integrals and then form the Lagrangian as follows:
\begin{equation}
\begin{split}
\mathcal{L}(\theta, z)= \int_{0}^{\infty} &\rme^{-\theta TB\mI (\mu (\theta, z)z)}f(z)dz\\&\hspace{1cm}+\lambda \Big(\int_{0}^{\infty}\mu (\theta, z)f(z)dz-\tSNR\Big).
\end{split}
\end{equation}
Above, $\lambda$ denotes the Lagrange multiplier. Setting the derivative of the Lagrangian with respect to $\mu (\theta, z)$ equal to zero, we obtain
\begin{equation}
\begin{split}
&\left. \frac{\partial \mathcal{L}(\mu(\theta,z),\lambda)}{\partial \mu(\theta,z)} \right |_{\mu(\theta,z)=\mu^*(\theta,z)} = 0
\\
&\Longrightarrow \Big(\lambda-\beta \rme^{-\theta TB \mI (\mu ^* (\theta, z)z)}\text{MMSE}(\mu ^* (\theta, z)z)z\Big)f(z)\!=\!0.
\end{split}
\end{equation}
Above, we have used the relation between the mutual information and MMSE given in (\ref{eq:first_der_mutual}) and defined $\beta=\theta TB \log_2 \rme$.  Let $\alpha=\frac{\lambda}{\beta}$.
Rearranging the above expression inside the parentheses, we obtain the equation in (\ref{eq:Ec_opt_equation}) where $\alpha$ can be found from the average power constraint given in (\ref{eq:alpha_constraint}). \hfill $\square$

Solving the equation in (\ref{eq:Ec_opt_equation}) does not result in a closed-form expression for $\mu ^* (\theta, z)$. We next show that the equation in (\ref{eq:Ec_opt_equation}) has at most one solution, denoted by $\mu ^* (\theta, z)$. Hence numerical root finding methods, e.g., bisection method, can efficiently determine $\mu ^* (\theta, z)$ \cite{boyd}.

\begin{Prop} The optimization problem in (\ref{eq:EC_opt}) has at most one solution.
\end{Prop}
\emph{Proof:} We first rewrite the equation in (\ref{eq:Ec_opt_equation}) by using the relation in (\ref{eq:first_der_mutual}) as follows:
\begin{align} \label{eq:g_func}
g\big((\mu(\theta, z)z)\big) = \rme^{-\theta TB \mI (\mu (\theta, z)z)}\dot{\mI} (\mu (\theta, z)z)z\log2  -\alpha.
\end{align}
Then, differentiating $g\big((\mu(\theta, z)z)\big)$ with respect to $(\mu(\theta, z)z)$ results in
\begin{equation}
\begin{split}
\dot{g}\big((\mu(\theta, z)z)\big)
=& \rme^{-\theta TB \mI (\mu (\theta, z)z)}z^2 \, \log2
\\
&\times
\Big(-\theta TB(\dot{\mI} (\mu (\theta, z)z))^2 +\ddot{\mI} (\mu (\theta, z)z)\Big).
\end{split}
\end{equation}
Since $\ddot{\mI} (\rho)<0$, the first derivative of $g\big((\mu(\theta, z)z)\big) $ is always negative, i.e., $\dot{g}\big((\mu(\theta, z)z)\big)<0$. Hence, using Rolle's theorem \cite{strang}, the equation in (\ref{eq:Ec_opt_equation}) cannot have more than one root. It is easily seen that when $\mu(\theta,z)=0$, $g\big((\mu(\theta, z)z)\big)= z-\alpha$, which is greater than $0$ when $z>\alpha$. As $\mu(\theta,z) \rightarrow \infty$, the first term in $(\ref{eq:g_func})$ is $0$ since $\dot{\mI} (\mu (\theta, z)z)=0$ by using the relation in (\ref{eq:first_der_mutual}). As a result, $g\big((\mu(\theta, z)z)\big)=-\alpha$, which is less than or equal to zero by definition of $\alpha$. Hence, we can conclude that there exists a unique optimal power policy for $z>\alpha$. \hfill $\square$

Therefore, there exists unique optimal power level denoted by $\mu ^*(\theta, z)$. \hfill $\square$

In Table \ref{table:algorithm}, the proposed power control algorithm that maximizes the effective capacity with an arbitrary input distribution subject to an average power constraint is summarized, where $\alpha$ in (\ref{eq:Ec_opt_equation}) is determined by using the projected subgradient method. In this method, $\alpha$ is updated iteratively according to the subgradient direction until convergence as follows:
\begin{align}
&\alpha^{(\!n+1\!)}\!\!=\!\!\Big[\alpha^{(\!n\!)}\!\!-\!\zeta\big( \tSNR\!-\!\E\{\mu ^* (\theta, z)\}\big)\Big]^+
\end{align}
where $[x]^+ =\max\{0,x\}$, $n$ is the iteration index and $\zeta$ is the step size. When $\zeta$ is chosen to be constant, it was shown that the subgradient method is guaranteed to converge to the optimal value within a small range \cite{boyd2}.
\begin{table}
\caption{} \label{table:algorithm}
\begin{algorithm}[H]
    \caption{Proposed power control algorithm for the effective capacity maximization with arbitrarily distributed inputs under an average power constraint}
    \begin{algorithmic}[1]
      \STATE Initialization: $\mu_{h}(\theta,z)=\mu_{h,\text{init}}$, $\mu_{l}(\theta,z)=\mu_{l,\text{init}}$, $\varepsilon > 0$, $\delta > 0$, $\zeta > 0$, $\alpha^{(0)}=\alpha_{\text{init}}$
      \REPEAT
      \STATE $n \leftarrow 0$
      \REPEAT
       \STATE update $\mu^*(\theta,z)=\frac{1}{2}(\mu_{h}(\theta,z)+\mu_{l}(\theta,z))$
	\STATE if $g(\mu^*(\theta,z))g(\mu_h(\theta,z)) <0$ (where $g(.)$ is defined in (\ref{eq:g_func})), then
        \STATE \hspace{1cm}$\mu_l(\theta,z) \leftarrow \mu^*(\theta,z)$
        \STATE else if $g(\mu^*(\theta,z))g(\mu_l(\theta,z)) <0$, then
        \STATE \hspace{1cm}$\mu_h(\theta,z) \leftarrow \mu^*(\theta,z)$
        \STATE end if
 \UNTIL{$|g(\mu^*(\theta,z))|< \varepsilon$}
\STATE update $\alpha$ using the projection subgradient method as follows
       \STATE $\alpha^{(n+1)}=\big[\alpha^{(n)}-\zeta(\tSNR-\E\{\mu^*(\theta,z)\})\big]^+$
\STATE $n \leftarrow n+1$
       \UNTIL{$|\alpha^{(n)}(\tSNR-\E\{\mu^*(\theta,z)\})| \le \delta$}
    \end{algorithmic}
  \end{algorithm}
\vspace{-0.9cm}
\end{table}
\begin{Rem} When the input signal is Gaussian, we have $\text{MMSE}(\rho)=\frac{1}{1+\rho}$ and $\mI (\rho)=\log_2(1+\rho)$. Substituting these expressions into (\ref{eq:Ec_opt_equation}), we can see that the optimal power control policy reduces to
\begin{align}
\label{eq:opt_power_Gaussian}
\mu_{\text{opt}}(\theta, z) = \begin{cases}
0 &z \le \alpha,\\
\frac{1}{\alpha^{\frac{1}{\beta+1}}z^{\frac{\beta}{\beta+1}}}-\frac{1}{z} &z> \alpha,
\end{cases}
\end{align}
which has exactly the same structure as given in \cite{tang}.
\end{Rem}

\section{Optimal Power Control in Asymptotic Cases} \label{sec:opt_power_limiting}
In this section, we analyze two limiting cases of the proposed optimal power control, in particular, when the system is subject to extremely stringent QoS constraints (i.e., as $\theta \rightarrow \infty$) and vanishing QoS constraints (i.e., as $\theta \rightarrow 0$), respectively.

\subsection{Optimal Power Control under Extremely Stringent QoS Constraints}
Asymptotically, when $\theta \rightarrow \infty$, the system is subject to increasingly stringent QoS constraints and hence it cannot tolerate any delay. In this case, the transmitter maintains a fixed transmission rate and the optimal power control for extremely stringent QoS constraints is known from \cite{tang} to be given by the total channel inversion scheme as follows:
\begin{align} \label{eq:channel_inversion}
\mu_{\text{opt}}(z)=\frac{\mathcal{C}}{z},
\end{align}
where the constant $\mathcal{C}$ can be found by satisfying the average transmit power constraint with equality. In particular,
\begin{align} \nonumber
\int_{0}^{\infty}\frac{\mathcal{C}}{z}f(z)&dz= \tSNR.
\end{align}
In Nakagami-$m$ fading channel, the channel power gain is distributed according to the Gamma distribution
\begin{align}
f(z)=\frac{z^{m-1}}{\Gamma (m)} \Big(\frac{m}{\Omega}\Big)^{m}\rme^{-\frac{m}{\Omega}z} \text{ for } m \geq 0.5,
\end{align}
where $m$ is the fading parameter, $\Omega$ is the average fading power and $\Gamma(x)$ is the Gamma function \cite[eq. 8.310.1]{gradshteyn}. In this case, $\mathcal{C}$ is given by
\begin{align}
\mathcal{C}&=\frac{\tSNR}{\E\Big\{\frac{1}{z}\Big\}}=\begin{cases}  \frac{\tSNR \Omega (m-1)}{m}& m > 1 \\ 0 &m \le 1 \end{cases}.
\end{align}
It should be noted that Nakagami-$m$ fading can model different fading conditions, e.g. including Rayleigh fading (i.e., $m=1$) and one-sided Gaussian fading (i.e., $m=0.5$) as special cases. Also, Nakagami-$m$ fading distribution is commonly used to characterize the received signal in urban radio \cite{suzuki} and indoor-mobile multipath propagation environments \cite{sheikh}.

\begin{Rem}
The power control policy under very stringent QoS constraints in (\ref{eq:channel_inversion}) is the same regardless of the signaling distribution while the effective capacity depends on the input distribution through mutual information expression. More specifically, with channel inversion power control policy, the mutual information becomes a constant, independent of the channel fading, i.e., $\mI (\mu (\theta, z)z) = \mI(\mathcal{C}) = \mI\left(\frac{\tSNR}{\E\left\{\frac{1}{z}\right\}}\right)$. Therefore, the effective capacity achieved with this policy can be expressed as
\begin{align*}
C_E(\tSNR) &= -\frac{1}{\theta TB}\log(\E\big\{\rme^{-\theta TB \,\, \mI (\mathcal{C})}\big\})
\\
&= -\frac{1}{\theta TB}\log(\rme^{-\theta TB \,\, \mI (\mathcal{C})})
\\
&= \mI (\mathcal{C}) =  \mI\left(\frac{\tSNR}{\E\left\{\frac{1}{z}\right\}}\right)
\end{align*}
which can be regarded as  the delay-limited rate achieved with a given input. For instance, with Gaussian signaling, we have the delay-limited capacity $\mI (\mathcal{C}) = \log_2 (1 + \mathcal{C}) = \log_2 \left( 1 + \frac{\tSNR}{\E\left\{\frac{1}{z}\right\}}\right)$ \cite{gursoy}.
\end{Rem}

\subsection{Optimal Power Control under vanishing QoS Constraints}
As $\theta \rightarrow 0$, QoS constraints eventually vanish, and hence the system can tolerate arbitrarily long delays. In this case, the effective capacity is equivalent to the achievable (mutual information) rate with finite discrete inputs. Subsequently, the optimization problem is expressed as
\begin{align}
\label{eq:rate_opt}
&\max_{
\substack{\mu(z)}} \E\{ \mI (\mu (\theta, z)) \}\\ \label{eq:avg_constraint}
&\hspace{-0.6cm}\text{subject to} \hspace{0.2cm}\E\{\mu (\theta, z)\}  \le \tSNR.
\end{align}
By following similar steps as in the proof of Theorem \ref{teo1}, the optimal power control policy is  given by
\begin{align} \label{eq:opt_power_noQoS}
\mu_{\text{opt}}(z)=\frac{1}{z}\text{MMSE}^{-1}\Big(\min \Big\{ 1,\frac{\eta}{ \log_2(e)z}\Big\}\Big).
\end{align}
Above, $\text{MMSE}^{-1}(.) \in [0,\infty )$ denotes the inverse MMSE function and the Lagrange multiplier, $\eta$ can be found by inserting the proposed power control into the power constraint in (\ref{eq:avg_constraint}) and satisfying this constraint with equality as follows:
\begin{align}
\int_{\frac{\eta}{\log_2(e)}}^{\infty}\mu_{\text{opt}}(z)f(z)dz = \tSNR.
\end{align}
\begin{Rem}
The power control policy in the absence of QoS constraints in (\ref{eq:opt_power_noQoS}) has the same structure of mercury/water-filling \cite{lozano}. It is seen that the power level depends on the input distribution through the expression of inverse MMSE.
\end{Rem}

\section{Low-Power Regime Analysis} \label{sec:low_power_analysis}
In this section, we study the performance in the low-power regime  achieved with arbitrary input distributions depending on the availability of CSI at the transmitter. In particular, we initially assume that the transmitter does not have the knowledge of the channel conditions and only the receiver has perfect CSI. In this setting, we consider constant-power transmissions and address the energy efficiency in the low-power regime via the first and second derivatives of the effective capacity. Subsequently, we assume both the transmitter and receiver have perfect CSI and characterize the optimal power control in this regime.

\subsection{Constant Power Transmissions} \label{subsec:constantpower}
Here, we assume that only the receiver has perfect CSI, and hence the signal is sent with constant power. In the low-power regime, EE can be characterized by the minimum energy per bit $\ben_{\rm{min}}$ and wideband slope $S_0$ \cite{verdu}. First, energy per bit is defined as
\begin{align}
\frac{E_{b}}{N_0} =\frac{\tSNR}{C_E(\tSNR)}.
\end{align}
Consequently, the minimum energy per bit required for reliable communication under QoS constraints is obtained from \cite{verdu}, \cite{gursoy}
\begin{equation}\label{eq:minbitenergy}
\ben_{\rm{min}}=\lim_{\tSNR\rightarrow 0}\frac{\tSNR}{C_E(\tSNR)}=\frac{1}{\dot{C}_E(0)},
\end{equation}
where $\dot{C}_E(0)$ denotes the first derivative of the effective capacity $C_E(\tSNR)$ with respect to $\tSNR$ in the limit as $\tSNR$ vanishes. Correspondingly, at $\ben_{\rm{min}}$, $S_0$ represents the linear growth of the spectral efficiency with respect to $\frac{E_{b}}{N_0}$ (in dB), which is obtained from \cite{verdu}, \cite{gursoy}
\begin{equation}\label{eq:widebandslope}
 S_0= \frac{-2(\dot{C}_E(0))^2}{\ddot{C}_E(0)}\log 2.
\end{equation}
Above, $\ddot{C}_E(0)$ denotes the second derivative of $C_E(\tSNR)$ with respect to $\tSNR$ in the limit as $\tSNR$ approaches zero. By using the minimum energy per bit in (\ref{eq:minbitenergy}) and wideband slope expression in (\ref{eq:widebandslope}), throughput can be approximated as a linear function of the energy per bit (in dB) as follows:
\begin{align} \label{eq:EC_low_power}
C_E=\frac{S_0}{10\log_{10}(2)}\bigg(\!\frac{E_{b}}{N_0}_{\text{dB}}\!-\ben_{\rm{min},\text{dB}}\bigg)\!+o\bigg(\frac{E_{b}}{N_0}_{\text{dB}}-\ben_{\rm{min},\text{dB}}\bigg),
\end{align}
where $\frac{E_{b}}{N_0}_{\text{dB}}=10\log_{10}\frac{E_{b}}{N_0}$ is the energy per bit in dB, and $o(\cdot)$ denotes the terms vanishing faster than the linear term.

We characterize these two important energy efficiency metrics in the low-power regime under QoS constraints in the following result.

\begin{Lem} The minimum energy per bit and wideband slope with arbitrary input distributions under QoS constraints for general fading distributions are given, respectively, by
\begin{align} \label{eq:min_energy_slope_arbitrary}
\ben_{\rm{min}}= \frac{\log 2}{\E\{z\}} \text{ and } S_0= \frac{2}{(-\ddot{\mI}(0)\log 2+\beta)\frac{\E\{z^2\}}{(\E\{z\})^2}-\beta},
\end{align}
where $\ddot{\mI}(0)$ denotes the second derivative of the mutual information evaluated at $\tSNR = 0$, and $\beta=\theta TB \log_2 \rme$.
\end{Lem}
\emph{Proof:} We first express the mutual information achieved with arbitrary input distributions in the low-power regime as follows:
\begin{align} \label{eq:mutual_low}
\mI (\tSNR z)= \tSNR z\log_2(e)+ \frac{\ddot{\mI} (0)}{2} \tSNR^2 z^2+ o(\tSNR^2).
\end{align}
Inserting the above expression into the effective capacity formulation, $C_E(\tSNR)$, given in (\ref{eq:EC_opt}) and evaluating the first and second derivatives of $C_E(\tSNR)$ with respect to $\tSNR$ at $\tSNR=0$ results in
\begin{align}
&\hspace{2cm}\dot{C}_E(0)= \frac{\E\{z\}}{\log 2} \\
\ddot{C}_E(0)&=(\ddot{\mI} (0)-\beta \log_2 \rme)\E\{z^2\}+ \beta \log_2\rme (\E\{z\})^2.
\end{align}
Further inserting the above expressions into those in (\ref{eq:minbitenergy}) and (\ref{eq:widebandslope}), the minimum energy per bit and wideband slope expressions in (\ref{eq:min_energy_slope_arbitrary}) are readily obtained. \hfill $\square$

From the above result, we immediately see that the same minimum energy per bit is achieved regardless of the signaling distribution and QoS constraints. On the other hand, the wideband slope depends on both the input distribution through $\ddot{\mI} (0)$, and the QoS exponent, $\theta$. More speficially, for quadrature symmetric constellations such as QPSK, $8$-PSK or $16$-QAM, we have $\ddot{\mI} (0) = -\log_2(e)$ while real-valued constellations such as BPSK and $m$-PAM lead to $\ddot{\mI} (0) = -2\log_2(e)$ \cite{lozano}. Hence, even though they have the same minimum energy per bit, quadrature symmetric constellations have higher wideband slopes compared to real-valued constellations, yielding higher EE.

It should also be noted that we obtain the low power behavior of the mutual information exhibited by the Gaussian input by setting $\ddot{\mI} (0) = -\log_2(e)$ in (\ref{eq:mutual_low}). Hence, substituting $\ddot{\mI} (0) = -\log_2(e)$ in (\ref{eq:min_energy_slope_arbitrary}), the minimum energy per bit and wideband slope expressions can be specialized to the case of Gaussian input, which leads the same formulations as in \cite{gursoy} under the assumption of perfect CSI only at the receiver.

\begin{Rem}
For a Nakagami-$m$ fading channel, $\E\{z\}=\Omega$ and $\E\{z^2\}=\Omega^2\big(1+\frac{1}{m}\big)$. Inserting these expressions into (\ref{eq:min_energy_slope_arbitrary}), the minimum energy per bit and wideband slope for a Nakagami-$m$ fading channel can be found, respectively, as
\begin{align} \label{eq:min_EbN0_S0_Nakagami}
\ben_{\rm{min}}= \frac{\log 2}{\Omega}, \text{ and } S_0=\frac{2}{-\big(1+\frac{1}{m}\big)\ddot{\mI} (0) \log2 + \frac{\beta}{m}}.
\end{align}
We note that while the minimum bit energy depends only on the average fading power, $\Omega$, the wideband slope is a function of Nakagami-$m$ fading parameter, input distribution and QoS exponent, $\theta$ (through the term $\beta=\theta TB \log_2 \rme$).
\end{Rem}

When there exists a dominant line of sight component along the propagation path, the Rician fading channel is an accurate model. This type of fading typically occurs in microcellular (e.g., suburban land-mobile radio communication) \cite{stewart} and picocellular environments (e.g., indoor communication) \cite{bultitude}. In this case, the pdf of the channel power gain is given by
\begin{align}
\small
\begin{split}
f(z)=\frac{(1+K)\rme^{-K}}{\Omega}\rme^{-\frac{(K+1)z}{\Omega}}I_0\bigg(2\sqrt{\frac{K(K+1)z}{\Omega}}\bigg) \text{ for } K, \Omega \geq 0,
\end{split}
\normalsize
\end{align}
where $K$ denotes the Rician $K$-factor and $I_0(x)$ represents the zero-th order modified Bessel function of the first kind \cite[eq. 8.405.1]{gradshteyn}.

\begin{Rem} By substituting $E\{z\}=\Omega$ and $\E\{z^2\}=\frac{(2+4K+K^2)\Omega}{(K+1)^2}$ into (\ref{eq:min_energy_slope_arbitrary}), we obtain the minimum energy per bit and wideband slope for the Rician fading channel as follows:
\begin{gather} \label{eq:Min_EbN_S0}
\begin{split}
&\ben_{\rm{min}} = \frac{\log(2)}{\Omega} \quad \text{ and }
\\
&S_0=\frac{2(K+1)^2}{-(2+4K+K^2)\ddot{\mI} (0)\log(2) + (2K+1)\beta}.
\end{split}
\end{gather}
It can be easily verified that the wideband slope is an increasing function of the Rician $K$-factor. Also, similar to the Nakagami-$m$ fading channel, the minimum energy per bit for the Rician fading channel depends only on the average fading power, $\Omega$.
\end{Rem}

\subsection{Optimal Power Control}
In this subsection, we assume that both the transmitter and receiver have perfect CSI. Below, we identify the optimal power control policy in the low-power regime.
\begin{Lem} \label{theo:optimalpowerlowSNRregime} The optimal power policy that maximizes the effective capacity with arbitrary input distributions in the low power regime is given by
\begin{align}\label{eq:ot_power_lowSNR}
\mu^*_{\text{opt}}(\theta,z)=\frac{z-\alpha}{\big(\beta-\ddot{\mI}(0)\big)z^2}.
\end{align}
\end{Lem}
\emph{Proof:} In the low power regime, MMSE behaves as \cite{lozano}
\begin{align}
\text{MMSE}(\rho) =1+\ddot{\mI}(0)\rho+\mathcal{O}(\rho^2)
\end{align}
which follows from the first-order Taylor series expansion of MMSE and the fact that the MMSE is proportional to the derivative of the mutual information \cite{guo}.
Incorporating the above approximation into (\ref{eq:Ec_opt_equation}), we have
\begin{equation}
\begin{split}
&\rme^{-\beta \int_{0}^{\mu(\theta,z)z}(1+\ddot{\mI}(0)\rho+\mathcal{O}(\rho^2))d\rho}
\\
&\times\Big(1+\ddot{\mI}(0)\mu(\theta,z)z+\mathcal{O}\big((\mu(\theta,z)z)^2\big)\Big)z=\alpha.
\end{split}
\end{equation}
Via the first-order Taylor expansion of the above equation, we obtain
\begin{align}
\Big(1-\big(\beta-\ddot{\mI}(0)\big)\mu(\theta,z)z+\mathcal{O}\big((\mu(\theta,z)z)^2\big)\Big)z=\alpha.
\end{align}
Solving the above equation provides the optimal power policy in (\ref{eq:ot_power_lowSNR}) where $\alpha$ is again found by satisfying the average power constraint as in (\ref{eq:alpha_constraint}).\hfill $\square$

For Nakagami-$m$ fading channel, $\alpha$ can be determined as the solution of
\begin{align}
\frac{-m^2\alpha \Gamma(m-2,\frac{m\alpha}{\Omega})+\Omega m\Gamma(m-1,\frac{m\alpha}{\Omega})}{\Omega^2\Gamma(m)}=\big(\beta-\ddot{\mI}(0)\big)\tSNR,
\end{align}
where $\Gamma(a,x)$ denotes the upper incomplete gamma function \cite[eq. 8.350.2]{gradshteyn}.

\section{Optimal Power Control under a Minimum EE Constraint}\label{sec:opt_power_EEmin}
In this section, we analyze the tradeoff between the EE and the effective capacity achieved with arbitrary input distributions by formulating the optimization problem to maximize the effective capacity subject to minimum EE and average transmit power constraints. More specifically, the optimization problem is expressed as
\begin{align}
\label{eq:EC_EEmin_opt}
C_E^{\text{opt}} (\tSNR)=&\max_{
\substack{\mu( \theta, z)}} -\frac{1}{\theta TB}\log\E\big\{\rme^{-\theta TB \mI (\mu (\theta, z)z)}\big\} \\
 \label{eq:EE_min_cons}\ &\hspace{-1cm}\text{subject to} \hspace{0.5cm} \frac{-\frac{1}{\theta TB}\log\E\big\{\rme^{-\theta TB \mI (\mu (\theta, z)z)}\big\}}{N_0B(\frac{1}{\xi}\E\{\mu (\theta,z)\}+P_{c_n})} \geq \text{EE}_{\text{min}}
\\  \label{eq:Pavg_constraint2} &\hspace{0.9cm}\E\{\mu (\theta, z)\}  \le \tSNR,
\end{align}
\normalsize
where $P_{c_n}$ represents the normalized circuit power, $\text{EE}_{\text{min}}$ denotes the minimum required EE, and $\xi$ is the power amplifier efficiency. In the following, we first derive the optimal power control subject to a minimum EE constraint in (\ref{eq:EE_min_cons}) and then address the average power constraint given in (\ref{eq:Pavg_constraint2}).

\begin{Lem} \label{teo1-ee}The optimal power control policy maximizing the effective capacity achieved with an arbitrarily distributed input subject to a minimum EE constraint is obtained as
\begin{align}
\label{eq:opt_power_EEmin}
\mu_{\text{opt}}(\theta, z) = \tilde{\mu} (\theta, z).
\end{align}
%
Above, $\tilde{\mu} (\theta, z)$ is the solution to the equation
\begin{align}
\rme^{-\theta TB \mI (\tilde{\mu} (\theta, z)z)}&\text{MMSE}(\tilde{\mu} (\theta, z)z)z \nonumber
\\
&=\frac{\nu \text{EE}_{\text{min}} N_0B \E\{\rme^{-\theta TB\mI (\tilde{\mu} (\theta, z)z)}\}}{\xi (1+\nu) \log_2(e)} \label{eq:Ec_opt_equation1}
\end{align}
where the Lagrange multiplier $\nu$ can be found by solving the equation below:
\begin{align}
\begin{split}
-\frac{1}{\theta TB}&\log\E\big\{\rme^{-\theta TB \mI (\mu_{\text{opt}} (\theta, z)z)}\big\}
\\
&-\text{EE}_{\text{min}}N_0B\Big(\frac{1}{\xi}\E\{\mu_{\text{opt}} (\theta,z)\}+P_{c_n}\Big)=0.
\end{split}
\end{align}
Consequently, the required SNR that satisfies the minimum EE is calculated as
\begin{align}\label{eq:Pavg_EEmin}
\tSNR^*=\E\{\mu_{\text{opt}}(\theta, z)\}.
\end{align}
\end{Lem}

\emph{Proof:} The objective function, $C_E(\tSNR)$ is concave in transmission power (as shown in the proof of Theorem \ref{teo1}) and total power consumption in the denominator of (\ref{eq:EE_min_cons}) is both affine and positive, hence the feasible set defined by (\ref{eq:EE_min_cons}), i.e., $S=\left\{\mu: C_E(\tSNR)-\text{EE}_{\text{min}}N_0B\Big(\frac{1}{\xi}\E\{\mu (\theta,z)\}+P_{c_n}\Big) \geq 0 \right\}$ is a convex set. Therefore, the Karush-Kuhn-Tucker conditions are sufficient and necessary to find the optimal solution. First, the minimum EE constraint in (\ref{eq:EE_min_cons}) can be rewritten as
\begin{align}
\begin{split}
-\frac{1}{\theta TB}&\log\E\big\{\rme^{-\theta TB \mI (\mu (\theta, z)z)}\big\}
\\
&-\text{EE}_{\text{min}}N_0B\Big(\frac{1}{\xi}\E\{\mu (\theta,z)\}+P_{c_n}\Big) \geq 0.
\end{split}
\normalsize
\end{align}
Let us define $\nu$ as the Lagrange multiplier associated with the minimum EE constraint. Then, the Lagrangian function is given by
\begin{align} \nonumber
&\mathcal{L}(P(g,h),\nu)=(1+\nu)\Big(-\frac{1}{\theta TB}\log\E\big\{\rme^{-\theta TB \mI (\mu (\theta, z)z)}\big\}\Big)\\&\hspace{2.8cm}-\nu \text{EE}_{\text{min}}N_0B\Big(\frac{1}{\xi}\E\{\mu (\theta,z)\}+P_{c_n}\Big).
\end{align}
Differentiating the above Lagrangian function with respect to $\mu(\theta,z)$ and and setting the derivative equal to zero, we obtain
\begin{align}
\small
\begin{split} \label{eq:Lagrangian_func_EEmin}
&\left. \frac{\partial \mathcal{L}(\mu(\theta,z),\nu)}{\partial \mu(\theta,z)} \right |_{\mu(\theta,z)=\tilde{\mu}(\theta,z)}
\\
&= (1+\nu)\frac{\log_2(e)\text{MMSE}(\tilde{\mu} (\theta, z)z)z\rme^{-\theta TB \mI (\tilde{\mu} (\theta, z)z)}}{\E\{\rme^{-\theta TB \mI (\tilde{\mu}(\theta, z)z)}\}}
\\
&\hspace{.5cm}-\nu \frac{\text{EE}_{\text{min}}}{\xi}N_0B=0.
\end{split}
\end{align}
\normalsize
Rearranging the terms in (\ref{eq:Lagrangian_func_EEmin}) leads to the desired result in (\ref{eq:Ec_opt_equation1}) and the Lagrange multiplier $\nu$ can be found by solving the equation in (\ref{eq:Ec_opt_equation1}) and then inserting the optimal power control into the minimum EE constraint. Consequently, the average transmission power is determined by substituting the optimal power control in (\ref{eq:opt_power_EEmin}) into (\ref{eq:Pavg_EEmin}). \hfill $\square$\

Now, we incorporate the average $\tSNR$ constraint in (\ref{eq:Pavg_constraint2}) into the proposed power control in (\ref{eq:opt_power_EEmin}). More specifically, if $\tSNR < \tSNR^*$ and the maximum EE subject to the average $\tSNR$ constraint in (\ref{eq:Pavg_constraint2}) is less than $\text{EE}_{\text{min}}$, then the optimization problem is not feasible and the power level is set to zero, i.e., $\mu^*(\theta,z) =0$. Otherwise the optimal power control is found considering the following two cases:
\begin{itemize}
\item If $\tSNR \geq \tSNR^*$, $\tSNR$ constraint is loose. In this case, the optimal power control is given by (\ref{eq:opt_power_EEmin}) where the minimum EE constraint is satisfied with equality.
\item If $\tSNR < \tSNR^*$ and the maximum EE subject to the average $\tSNR$ constraint in (\ref{eq:Pavg_constraint2}) is greater than $\text{EE}_{\text{min}}$, the minimum EE constraint does not have any effect on the maximum effective capacity. In this case, the optimal power control is determined by (\ref{eq:opt_power}) where the average $\tSNR$ constraint is satisfied with equality.
\end{itemize}

\begin{Rem} Inserting $\text{MMSE}(\rho)=\frac{1}{1+\rho}$ and $\mI (\rho)=\log_2(1+\rho)$ into (\ref{eq:Ec_opt_equation1}), the optimal power control scheme for Gaussian distributed signal becomes
\begin{align}
\mu^*(\theta,z)=\begin{cases} 0 &z \le \gamma_1 \\
\frac{1}{\gamma_1^{\frac{1}{1+\beta}}z^{\frac{\beta}{1+\beta}}}-\frac{1}{z} &z > \gamma_1
\end{cases},
\end{align}
which is in agreement with the result obtained in \cite{musavian}. Above, $\gamma_1=\frac{\nu \text{EE}_{\text{min}} N_0B \E\{\rme^{-\theta TB\mI (\mu (\theta, z)z)}\}}{\xi (1+\nu) \log_2(e)}$ is the scaled Lagrange multiplier, which can be found by inserting the above power control into (\ref{eq:Ec_opt_equation1}) and solving the corresponding equation for $\gamma_1$.
\end{Rem}

\section{Numerical Results} \label{sec:num_results}
\begin{figure*}[ht]
\centering
\begin{subfigure}[b]{0.42\textwidth}
\centering
\includegraphics[width=\textwidth]{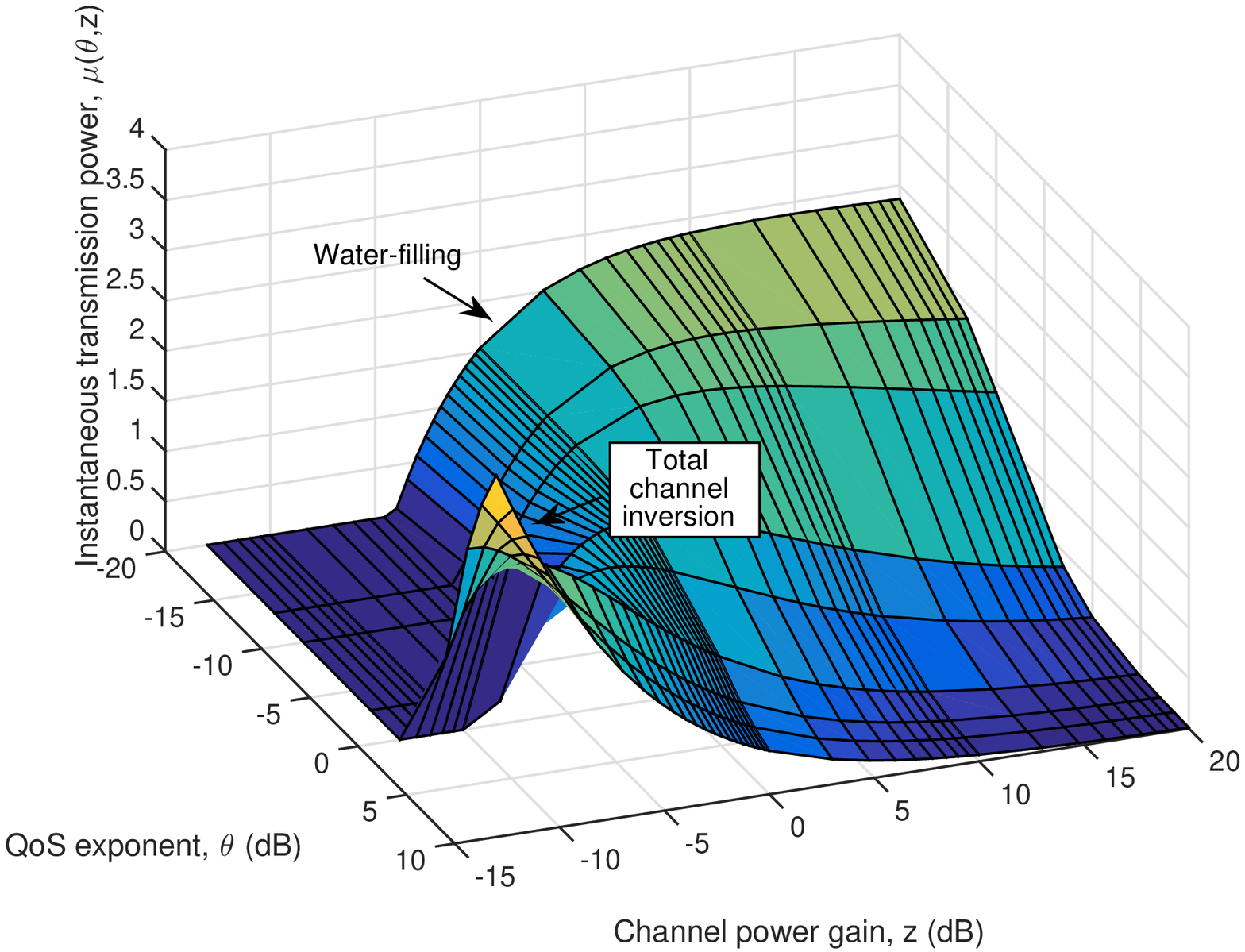}
\caption{}
\end{subfigure}
\begin{subfigure}[b]{0.42\textwidth}
\centering
\includegraphics[width=\textwidth]{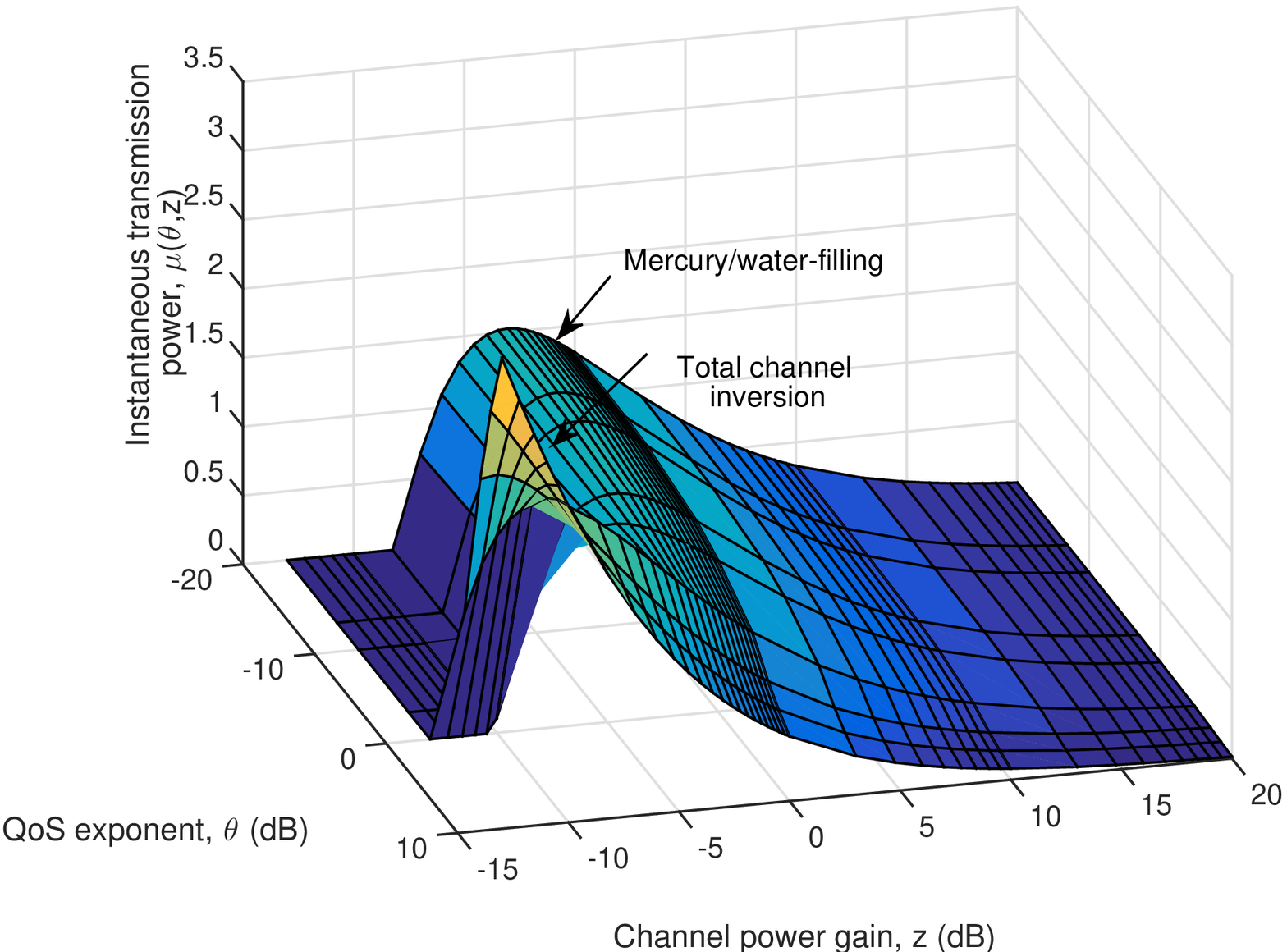}
\caption{}
\end{subfigure}
\caption{The instantaneous transmission power as a function of channel power gain, $z$ and QoS exponent, $\theta$ for (a) Gaussian input; (b) BPSK input .} \label{fig:powers_Gaussian_BPSK}
\end{figure*}
In this section, we present numerical results to illustrate the proposed optimal power control policies and the corresponding performance levels. Unless mentioned explicitly, we consider Nakagami-$m$ fading channel with $m=1$ (which corresponds to Rayleigh fading) in the simulations, and it is assumed that $TB=1$, $\Omega=1$ and average transmit power constraint, $\bar{P}=0$ dB. In the iterations, step size $\zeta$ is chosen as $0.1$, $\varepsilon$ and $\delta$ are set to $10^{-5}$.

In Fig.~\ref{fig:powers_Gaussian_BPSK}, we plot the instantaneous power level as a function of the channel power gain $z$ and the QoS exponent $\theta$ for both Gaussian and BPSK signals. As $\theta$ decreases, QoS constraint becomes looser. In this case, the power control for BPSK input has the structure of mercury/water-filling policy. In particular, the power is allocated to the better channel up to capacity saturation and then extra power is assigned to the worse channel. When the input is Gaussian, the power adaptation policy becomes the water-filling scheme, with which more power is assigned to the better channel opportunistically, deviating from the mercury/water-filling policy. When $\theta$ increases and hence stricter QoS constraints are imposed, the optimal power control policy becomes channel inversion for both inputs.

\begin{figure}[htb]
\centering
\includegraphics[width=0.5\textwidth]{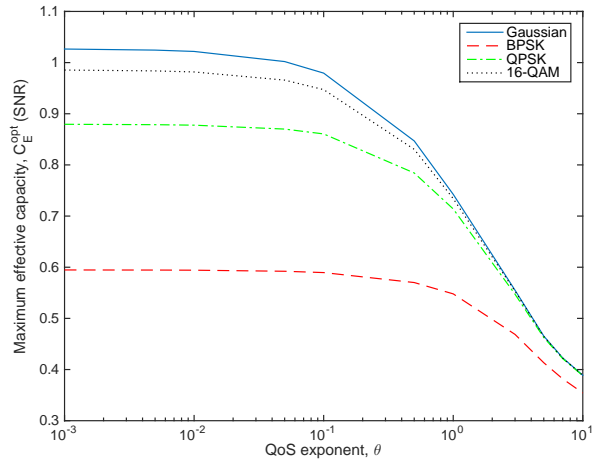}
\caption{Maximum effective capacity $C_E^{\text{opt}}(\tSNR)$ vs. QoS exponent $\theta$ for Gaussian, BPSK, QPSK and 16-QAM inputs.}
\label{fig:Ec_theta_inputs}
\end{figure}

In Fig.~\ref{fig:Ec_theta_inputs}, we display the effective capacity $C_E^{\text{opt}}(\tSNR)$ as a function of the QoS exponent $\theta$ for Gaussian, BPSK, QPSK and 16-QAM inputs with $\tSNR = 0$dB. It is observed that as $\theta$ increases, the effective capacity for all inputs decreases since the transmitter is subject to more stringent QoS constraints, which results in lower arrival rates hence lower effective capacity. It is also seen that Gaussian inputs always achieve higher effective capacity. For large $\theta$ values, Gaussian input and QPSK exhibit nearly the same performance. Therefore, under strict QoS constraints, QPSK can be efficiently used in practical systems rather than the Gaussian input which is difficult to implement.

\begin{figure}[htb]
\centering
\includegraphics[width=0.5\textwidth]{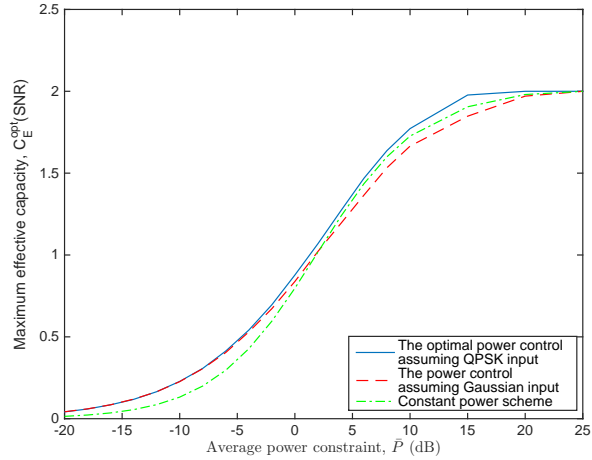}
\caption{Maximum effective capacity $C_E^{\text{opt}}(\tSNR)$ vs. average transmit power constraint $\bar{P}$ for QPSK input.}
\label{fig:Ec_Pavg_QPSK}
\end{figure}

In Fig.~\ref{fig:Ec_Pavg_QPSK}, we plot maximum effective capacity $C_E^{\text{opt}}(\tSNR)$ as a function of average transmit power constraint $\bar{P}$ for QPSK input (i.e., in all the curves we assume that QPSK signaling is employed). QoS exponent $\theta$ is set to $0.1$. We compare the performances of the constant-power scheme, power control assuming Gaussian input and the optimal power control assuming QPSK input. It is observed that as $\bar{P}$ increases, the effective capacity increases and then saturates due to the fact that the input is generated  from a finite discrete modulation. It is seen that the power control considering the true input distribution, in this case QPSK, achieves the highest effective capacity since the power control assuming Gaussian input is not the optimal policy for the QPSK input, and constant-power transmission strategy does not take advantage of favorable channel conditions. In addition, the performance gap between the optimal power control considering the discrete constellation and power control assuming Gaussian input increases at moderate SNR levels. Note that as shown in Theorem \ref{theo:optimalpowerlowSNRregime} with the expression in (\ref{eq:ot_power_lowSNR}), the optimal power control policy in the low power regime depends on the type of input via $\ddot{\mI}(0)$. As remarked in Section \ref{subsec:constantpower}, we have $\ddot{\mI}(0) = -\log_2(e)$ for both QPSK and Gaussian inputs. Therefore, at low power levels, we have the same power control policy regardless of whether it is designed for the QPSK input or the Gaussian input, and consequently the same effective capacity values are initially attained by the two power control policies in the low power regime as observed in Fig.~\ref{fig:Ec_Pavg_QPSK}. However, these two power control policies are no longer similar as power levels increase, leading to the observed performance gap at moderate SNR/power levels. At the other extreme, when the transmit power is sufficiently high, the throughput of QPSK saturates at 2 bits/symbol and expectedly, all curves eventually start converging.

\begin{figure}[htb]
\centering
\includegraphics[width=0.5\textwidth]{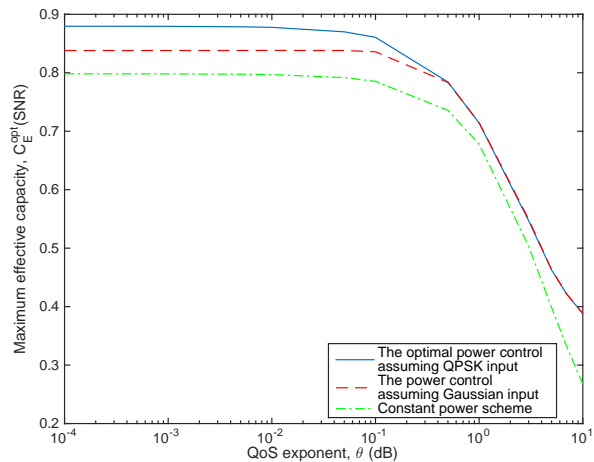}
\caption{Maximum effective capacity $C_E^{\text{opt}}(\tSNR)$ vs. QoS exponent $\theta$ for QPSK input.}
\label{fig:Ec_theta_QPSK}
\end{figure}

\begin{figure*}[ht]
\centering
\begin{subfigure}[b]{0.495\textwidth}
\centering
\includegraphics[width=\textwidth]{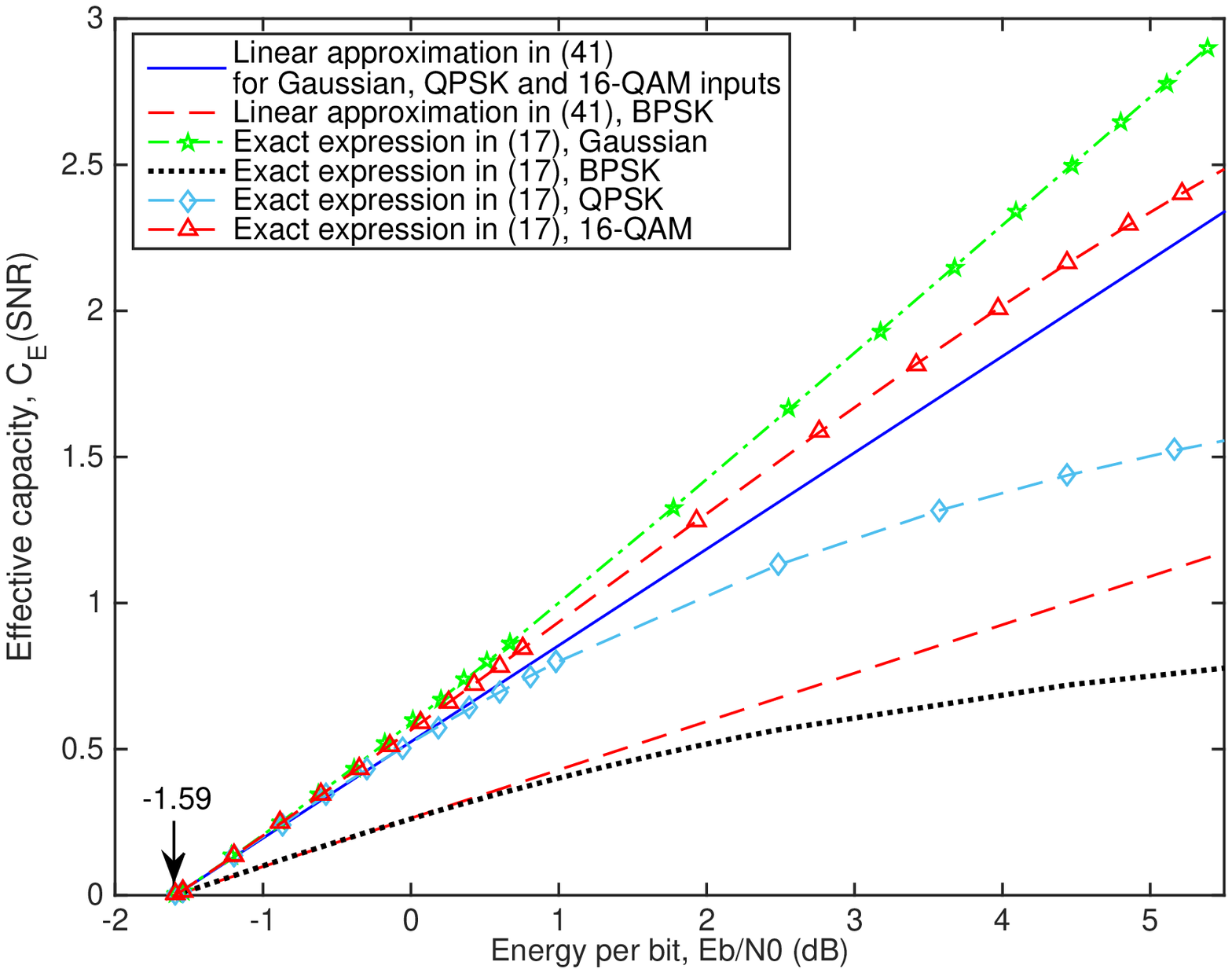}
\caption{}
\end{subfigure}
\begin{subfigure}[b]{0.495\textwidth}
\centering
\includegraphics[width=\textwidth]{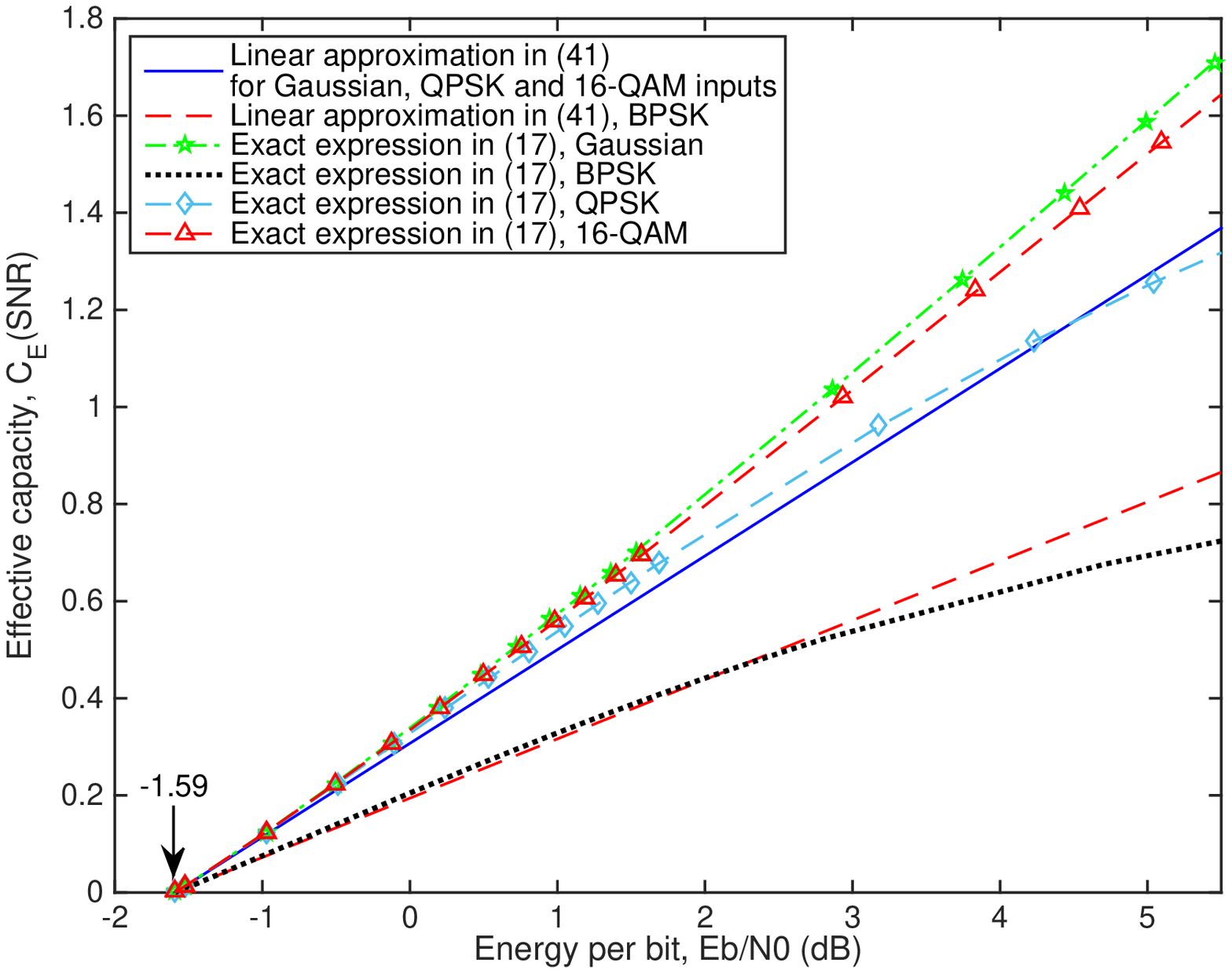}
\caption{}
\end{subfigure}
\caption{Effective capacity vs. energy per bit, $\frac{E_{b}}{N_0}_{\text{dB}}$ for Gaussian, BPSK, QPSK and 16-QAM inputs (a) $\theta=0.01$ and (b) $\theta=1$.}
\label{fig:Ec_EbN0}
\end{figure*}

In Fig.~\ref{fig:Ec_theta_QPSK}, maximum effective capacity $C_E^{\text{opt}}(\tSNR)$ as a function of the QoS exponent $\theta$ for QPSK input is illustrated when with $\tSNR = 0$dB. We again consider the constant-power scheme, power control assuming Gaussian input and the optimal power control assuming QPSK input. The constant-power scheme has the worst performance with the lowest effective capacity for all values of $\theta$. It is also interesting to note that the performance gap between the power control policies assuming Gaussian input and QPSK input is initially large for small values of $\theta$, and decreases as $\theta$ increases. This is mainly due to the fact that for higher values of $\theta$, the power control scheme does not depend on the input distribution and becomes total channel inversion.

\begin{figure}[htb]
\centering
\includegraphics[width=0.5\textwidth]{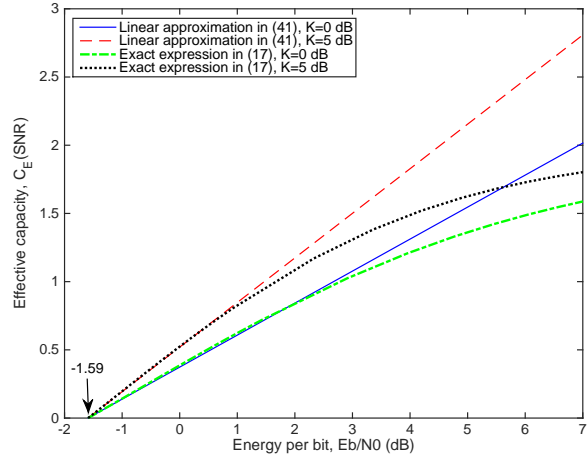}
\caption{Effective capacity vs. energy per bit, $\frac{E_{b}}{N_0}_{\text{dB}}$ for QPSK input in Rician fading channel.}
\label{fig:Ec_EbN0_K}
\end{figure}

In Fig.~\ref{fig:Ec_EbN0}, we plot the effective capacity as a function of energy per bit $\frac{E_{b}}{N_0}_{\text{dB}}$ for constant-power transmission when $\theta=0.01$ and $\theta=1$. We compare the performances of Gaussian, BPSK, QPSK and 16-QAM inputs in the low power regime by analyzing the minimum energy per bit and wideband slope values. It is observed that all inputs achieve the same minimum energy per bit of $-1.59$ dB while the wideband slope for BPSK is smaller than those of Gaussian, QPSK and 16-QAM inputs, which indicates lower EE for BPSK. Gaussian input achieves the highest EE among the inputs. We also consider the linear approximation for the effective capacity in the low power regime given in (\ref{eq:EC_low_power}) and the exact analytical effective capacity expression in (\ref{eq:effective_capacity}). It is seen that the linear approximation for all inputs is tight at low SNR values or equivalently low values of $E_b/N_0$ (dB). Additionally, when we compare Fig. 5a with Fig. 5b, we readily observe that the minimum energy per bit remains the same as QoS exponent $\theta$ changes from $0.01$ to $1$. On the other hand, wideband slope decreases with increasing $\theta$, which confirms the result in (\ref{eq:min_EbN0_S0_Nakagami}).

In Fig.~\ref{fig:Ec_EbN0_K}, we display effective capacity as a function of energy per bit, $\frac{E_{b}}{N_0}_{\text{dB}}$ for QPSK input. We consider Rician fading channel with different values of Rician $K$-factor (i.e., $K=0$ dB and $K=5$ dB). It is again observed that the linear approximation for the effective capacity in (\ref{eq:EC_low_power}) and the exact analytical effective capacity expression in (\ref{eq:effective_capacity}) matches well at low SNR values. Minimum energy per bit does not get affected by the Rician $K$-factor. However, as $K$ increases, EE increases as evidenced by the increased wideband slope. This observation is in agreement with the minimum energy per bit and wideband slope expressions in (\ref{eq:Min_EbN_S0}).

\begin{figure}[htb]
\centering
\includegraphics[width=0.5\textwidth]{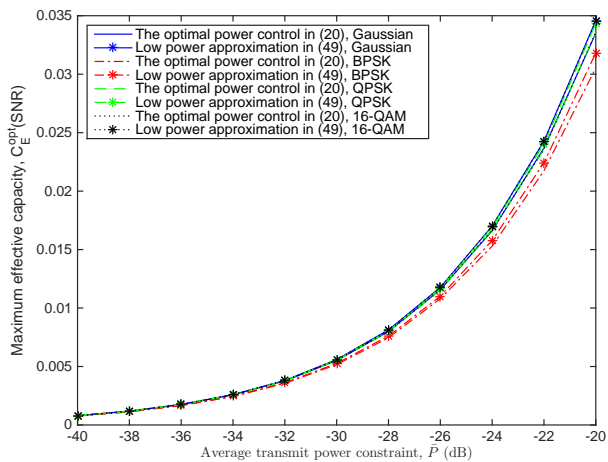}
\caption{Effective capacity vs. average transmit power constraint $\bar{P}$ for Gaussian, BPSK, QPSK and 16-QAM inputs.}
\label{fig:Ec_EbN0_power_control}
\end{figure}

In Fig.~\ref{fig:Ec_EbN0_power_control}, we plot effective capacity as a function of the average transmit power constraint $\bar{P}$ for Gaussian, BPSK, QPSK and 16-QAM inputs. We consider the proposed optimal power control in (\ref{eq:opt_power}) and low-power approximation for the power control in (\ref{eq:ot_power_lowSNR}). The figure validates the accuracy of the approximation at low power levels. Also, decreasing $\bar{P}$ leads to lower effective capacity for all inputs.

\begin{figure}[htb]
\centering
\includegraphics[width=0.5\textwidth]{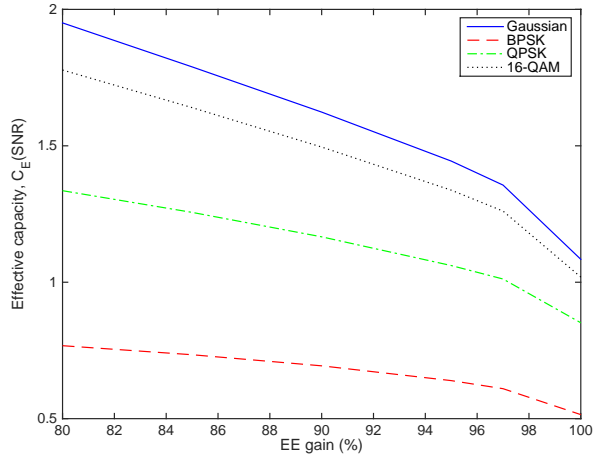}
\caption{Effective capacity gain vs. EE gain for Gaussian, BPSK, QPSK and 16-QAM inputs.}
\label{fig:Ec_EE_gain}
\end{figure}

Finally, we analyze the tradeoff between effective capacity and EE. In particular, we display effective capacity as a function of EE gain ($\%$) for Gaussian, BPSK, QPSK and 16-QAM inputs in Fig.~\ref{fig:Ec_EE_gain}. We assume that QoS exponent $\theta=0.1$ and average transmit power constraint $\bar{P}=6$ dB. The EE gain is determined as the ratio of the minimum required EE denoted by $\text{EE}_{\text{min}}$ to the maximum achievable EE. It is seen that the effective capacity decreases with increasing EE gain for all inputs, indicating that gains in energy efficiency is obtained at the expense of lower throughput. Again, Gaussian input achieves the highest effective capacity.

\section{Conclusion}\label{sec:conc}
In this paper, we have derived the optimal power control policies in wireless fading channels with arbitrary input distributions under QoS constraints by employing the effective capacity as the throughput metric. We have proposed a low-complexity optimal power control algorithm. We have analyzed two limiting cases of the optimal power control. In particular, when QoS constraints vanish, the optimal power allocation strategy converges to mercury/water-filling for finite discrete inputs and water-filling for Gaussian input, respectively. When QoS constraints are extremely stringent, the power control becomes the total channel inversion and no longer depends on the input distribution. It is observed that Gaussian input achieves the highest effective capacity among the inputs. Subsequently, we have analyzed the performance with arbitrary signal constellations at low spectral efficiencies by characterizing the minimum energy per bit and wideband slope for general fading distributions. The results are specialized to Nakagami-$m$ and Rician fading channels. We have shown that while the minimum energy per bit does not get affected by the input distribution, the wideband slope depends on both the QoS exponent, the input distribution and the fading parameter. We have determined the optimal power control policy in the low-power regime. The accuracy of the proposed power control is validated through numerical results. Finally, we have studied the effective capacity and EE tradeoff for arbitrary input signaling. In particular, we have solved the optimization problem to maximize the effective capacity achieved with arbitrarily distributed inputs subject to constraints on the minimum required EE and average power constraint.

\end{spacing}

\end{document}